\documentclass[twocolumn, amssymb, amsmath, aps, superscriptaddress, showpacs, footinbib,  prb]{revtex4}
\usepackage{graphicx}

\newcommand{\eff}{\text{eff}}
\newcommand{\AFM}{\text{AFM}}
\newcommand{\FM}{\text{FM}}
\newcommand{\LMTO}{\text{LMTO}}

\begin{document}

\title{Uniform spin chain physics arising from NCN bridges in CuNCN:\\ surprises on the way from copper oxides to their nitride analogs}
\author{Alexander A. Tsirlin}
\email{altsirlin@gmail.com}
\author{Helge Rosner}
\email{Helge.Rosner@cpfs.mpg.de}
\affiliation{Max Planck Institute for Chemical Physics of Solids, N\"{o}thnitzer
Str. 40, 01187 Dresden, Germany}

\begin{abstract}
We report on the unexpected uniform spin chain physics in CuNCN, the insulating nitride analog of copper oxides. Based on full-potential band structure calculations, we derive the relevant microscopic parameters, estimate individual exchange couplings, and establish a realistic spin model of this compound. The structure of CuNCN contains chains of edge-sharing CuN$_4$ squares. As a surprise, in contrast to analogous [CuO$_2$] chains in "edge-sharing" cuprates, the leading magnetic interactions $J\simeq 2500$~K run perpendicular to the structural [CuN$_2$] chains via bridging NCN groups. The resulting spin model of a uniform chain is in agreement with the experimentally observed temperature-independent magnetic susceptibility below 300~K. The nearest-neighbor and next-nearest-neighbor interactions along the structural [CuN$_2$] chains are $J_1\simeq -500$~K and $J_2\simeq 100$~K, respectively. Despite the frustrating nature of $J_1$ and $J_2$, we assign the susceptibility anomaly at 70~K to long-range magnetic ordering, which is likely collinear with antiparallel and parallel arrangement of spins along the $c$ and $a$ directions, respectively. The pronounced one-dimensionality of the spin system should lead to a reduction in the ordered moment and to a suppression of the transition anomaly in the specific heat, thus impeding the experimental observation of the long-range ordering. Our results suggest CuNCN as a promising material for ballistic heat transport within spin chains, while the sizable bandwidth $W\simeq 3$~eV may lead to a metal-insulator transition and other exotic properties under high pressure.
\end{abstract}

\maketitle

\section{Introduction}
One-dimensional (1D) electronic systems are in the focus of experimental and theoretical research due to the exotic properties that emerge in low-dimensional models and can be observed in real systems.\cite{giamarchi,quantum-wires} Copper(II) oxides with chain-like structures are one of the best playgrounds for studying 1D physics of localized electrons, because the strong Coulomb repulsion in Cu $d$ shell and the half-filling regime lead to insulating behavior, while the rich crystal chemistry of copper oxides allows to vary relevant microscopic parameters. Copper oxides can be properly described within simple or extended Hubbard models.\cite{emery} Subsequently, the low-energy properties of undoped (Cu$^{+2}$-containing) systems are easily mapped onto the spin-only Heisenberg model.\cite{models-1990,helge-m2cuo3}

The two main scenarios of Cu-based 1D systems are the so-called "corner-sharing" and "edge-sharing" chains that correspond to corner or edge connections between adjacent CuO$_4$ plaquettes within the chain. In "corner-sharing" chains, the Cu--O--Cu angle is usually close to $180^{\circ}$ and allows for the strong superexchange between neighboring Cu atoms. This type of the chain structure is found in Sr$_2$CuO$_3$, the archetypal material for the Heisenberg model of uniform spin-$\frac12$ chain.\cite{m2cuo3-neutrons,sr2cuo3-properties} This model allows for ballistic heat transport mediated by spin excitations.\cite{heat-transport-1,heat-transport-2} Indeed, Sr$_2$CuO$_3$ and other spin-chain materials reveal surprisingly high thermal conductivities along the direction of spin chains.\cite{sologubenko2000}

The case of the "edge-sharing" chains is quite different. Copper atoms are located closer to each other, and two competing magnetic interactions emerge. The nearest-neighbor (NN) coupling $J_1$ is usually ferromagnetic (FM), because the Cu--O--Cu angle is close to $90^{\circ}$. In contrast, the next-nearest-neighbor (NNN) coupling $J_2$ is antiferromagnetic (AFM) and corresponds to Cu--O--O--Cu superexchange. The absolute values of $J_1$ and $J_2$ are usually comparable. This leads to strong magnetic frustration. For $J_2/J_1<-0.25$, the energy of the system can be minimized in the spiral state, where adjacent spins are turned for a constant angle with respect to each other. Although the Mermin-Wagner theorem states the lack of the long-range ordering in a purely 1D system above zero temperature, non-negligible interchain couplings stabilize the finite-temperature spiral order in real materials. This type of the ordering was recently observed in a number of "edge-sharing" compounds, such as LiCu$_2$O$_2$,\cite{licu2o2-2004,masuda2004,masuda2005} LiCuVO$_4$,\cite{gibson2004,enderle2005} and Li$_2$CuZrO$_4$.\cite{drechsler2007,li2cuzro4-neutrons}

A spiral magnetic ground state discloses further unusual phenomena. Below the ordering temperature, LiCu$_2$O$_2$ and LiCuVO$_4$ show unconventional ferroelectricity which is strongly coupled to the magnetic field.\cite{park2007,licuvo4-ferroelectricity} However, the origin of this behavior remains controversial. The ferroelectricity can be explained either as a purely electronic effect (the tendency to release the frustration)\cite{moskvin2008-2} or as a result of Li/Cu antisite disorder.\cite{moskvin2008,moskvin2009} To get further insight into the ferroelectricity of Cu-based spin-chain materials, one has to study other compounds showing the spiral ground state, e.g., with different ligands. The size, charge, and chemical nature of the ligand control the relevant microscopic parameters: the hoppings and the magnitude of electronic correlations (the Coulomb repulsion in Cu $d$ shell, which is partially screened by the ligand orbitals). The search for new materials led to the recent study of CuCl$_2$ with chains of edge-sharing CuCl$_4$ squares.\cite{schmitt2009,cucl2-2009} Low-temperature neutron diffraction evidences its spiral magnetic structure\cite{cucl2-2009} and calls for the further study of the possible low-temperature ferroelectricity in this compound. Another suitable ligand is nitrogen. However, nitride compounds of Cu$^{+2}$ are scarce, because the nitride ligand usually stabilizes low oxidation states of transition metals -- e.g., +1 for Cu in the simple copper nitride Cu$_3$N.

\begin{figure}
\includegraphics{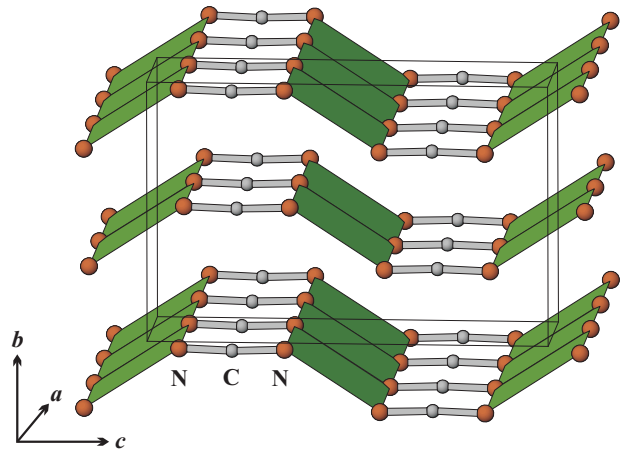}
\caption{\label{structure}
(Color online) Crystal structure of CuNCN showing chains of edge-sharing CuN$_4$ plaquettes. The chains are linked into layers via NCN groups. The Cu atoms are located in the centers of the green CuN$_4$ plaquettes.
}
\end{figure}

Recently, the nitride environment of Cu$^{+2}$ was achieved in copper cyanodiimide CuNCN.\cite{cuncn-2005} The crystal structure of this compound reveals chains of edge-sharing CuN$_4$ plaquettes arranged along the $a$ axis. The nearly linear NCN groups couple the chains in the $ac$ plane, while the resulting layers are stacked along the $b$ axis (Fig.~\ref{structure}). CuNCN has black color. Resistivity measurements on polycrystalline samples evidence semiconducting behavior with the activation energy of 0.1~eV. Below 300~K, the magnetic susceptibility of CuNCN is nearly temperature-independent. Additionally, a susceptibility anomaly is observed at 70~K, whereas no respective anomaly is seen in the specific heat. The low-temperature neutron diffraction does not manifest long-range magnetic ordering.\cite{cuncn-2008}

Band structure calculations suggested a spin model of an anisotropic triangular lattice with leading exchange couplings in the $ab$ plane.\cite{cuncn-2008} However, the relevant microscopic parameters of the Hubbard model were not derived. Additionally, the proposed spin model seems to be counter-intuitive from chemical point of view, because the strong interaction along the $b$ direction ($J_b$ of about 800~K according to Ref.~\onlinecite{cuncn-2008}, see Fig.~\ref{interactions} for notation) corresponds to the rather long Cu--Cu distance of 3.43~\r A and lacks any obvious superexchange pathway (see Fig.~\ref{structure}). The reference to oxide materials with "edge-sharing" chains\cite{licu2o2-2004,masuda2005,enderle2005,drechsler2007} suggests that the half-filled Cu $d$ orbital lies in the CuN$_4$ plane, while four other $d$ orbitals are fully occupied and do not take part in the magnetic interactions. Then, the interlayer couplings should be negligible, whereas strong exchange couplings should run along the structural chains and show magnetic frustration. 

In the following, we perform accurate full-potential band structure calculations that evaluate the relevant microscopic parameters and suggest a realistic microscopic model for CuNCN. We show that the "on-site" physics of this nitride material indeed resembles copper oxides, the only difference being a reduction in the on-site Coulomb repulsion due to the stronger screening caused by nitrogen as a ligand. However, the "inter-site" physics is dramatically changed, because NCN groups mediate unexpectedly strong superexchange couplings. Then, the magnetic behavior is 1D. However, the spin chains do not match the structural chains but rather run perpendicular to them. We carefully compare our results to the available experimental data, analyze similarities and differences between CuNCN and copper oxides, and provide an outlook for further experimental studies.

\section{Methods}
\label{methods}
Our microscopic model is based on band structure calculations performed within local density approximation (LDA) of density functional theory. We use the exchange-correlation potential by Perdew and Wang\cite{perdew} and the full-potential band structure code with a basis set of atomic-like local orbitals (FPLO8.50-32).\cite{fplo} To check the robustness of the results, we performed a number of comparative calculations applying the generalized gradient approximation (GGA) with the Perdew-Burke-Ernzerhof exchange-correlation potential.\cite{pbe} In order to compare our results to the previous study (Ref.~\onlinecite{cuncn-2008}), we also repeated several calculations using the projector augmented wave (PAW) method\cite{paw} implemented in the Vienna ab-initio simulation package (VASP).\cite{vasp}

For the LDA calculation, the proper partitioning of the reciprocal space is achieved by the $k$ mesh with 2041 points in the irreducible part of the first Brillouin zone. The convergence with respect to the $k$ mesh was carefully checked. We used two sets of structural parameters derived from the refinement of the room-temperature x-ray data\cite{cuncn-2005} and the low-temperature neutron data.\cite{cuncn-2008} The two sets of atomic coordinates yield essentially similar results, hence one should not expect any pronounced temperature-dependent changes in the electronic structure of CuNCN. 

While LDA provides a simple picture of electronic states and yields input parameters for the free-electron part of the Hubbard hamiltonian, it misses the strong electronic correlations in the $d$ shell and fails to describe the correlation-induced band gap in many transition metal compounds. To account for the correlation effects, one has to include them either on the model level (on top of self-consistent band structure calculations) or into the self-consistent procedure itself. For the model treatment of correlations, we fit relevant LDA bands with a tight-binding (TB) model and include the resulting hopping parameters into a one-orbital Hubbard model with the effective on-site Coulomb repulsion $U_{\eff}$. Then, the Hubbard model can be solved using numerical simulations. For the low-lying (spin) excitations and for the half-filling regime, we can reduce the hamiltonian to a Heisenberg model and calculate the exchange integrals via the simple expression of second-order perturbation theory. Further details of the procedure are given in Sec.~\ref{results}.

To include electronic correlations into the self-consistent procedure, one usually applies the local spin-density approximation (LSDA)+$U$ approach that treats correlations in a mean-field way. LSDA+$U$ gives reliable energies for different types of spin order, and these energies can be further mapped onto the classical Heisenberg model. This approach has already been utilized in Ref.~\onlinecite{cuncn-2008}. However, the authors of Ref.~\onlinecite{cuncn-2008} performed calculations for a relatively small supercell which allowed to access few exchange couplings only. In the following, we will show that further, long-range couplings should also be included in the model. To evaluate them, we use two 32-atom supercells: $2\mathbf{a}\times \mathbf{b}\times \mathbf{c}$ and $4\mathbf{a}\times\frac12(\mathbf a+\mathbf b)\times\mathbf c$, where $\mathbf a$, $\mathbf b$, and $\mathbf c$ are the translation vectors of the crystallographic $C$-centered unit cell of CuNCN. The $k$ mesh included 256 and 192 points, respectively. The resulting total energies allow to estimate all the couplings evidenced by the LDA-based analysis. The use of the two supercells has an additional advantage, because short-range couplings are independently estimated in each of the supercells. Then, the comparison of the resulting numbers suggests a natural error bar for the computational results. In contrast to other copper compounds (e.g., Ref.~\onlinecite{cucl}), the two supercells yield slightly different estimates of the exchange couplings. This issue will be further discussed in Sec.~\ref{results}.

The electronic correlations are parametrized by the effective on-site Coulomb repulsion potential $U_{\eff}$ of the Hubbard model or by the $U_d$ (repulsion) and $J_d$ (exchange) parameters of LSDA+$U$. To get an idea about the values of these parameters, we use the constrained LDA approach\cite{constrained} implemented in the TB-LMTO-ASA (tight-binding linearized muffin-tin orbitals in atomic spheres approximation) code.\cite{lmto} Under the constraint of zero hoppings between correlated and uncorrelated orbitals, we calculate the energy of the correlated ($3d$) orbital for several fixed occupancies. This yields the estimates of $U_d^{\LMTO}=6.6$~eV and $J_d^{\LMTO}=1.0$~eV for CuNCN. These numbers can be further used for LSDA+$U$ calculations within LMTO-ASA. However, the transfer to other codes or models requires a more careful consideration of the meaning regarding $U_d^{\LMTO}$ and $J_d^{\LMTO}$.

The constrained LDA procedure yields the $U_d$ and $J_d$ parameters which are relevant for the specific (muffin-tin) orbitals of the LMTO code. In other codes, these parameters are applied to different $d$ functions, hence their values should also be different. In the Hubbard model, one treats correlated bands rather than atomic orbitals. In the case of CuNCN, nearly half of the correlated bands originate from uncorrelated nitrogen states (see Sec.~\ref{results}), hence a strong screening is expected, and the $U_{\eff}$ value should be reduced compared to $U_d$ and $J_d$. Thus, the $U_d^{\LMTO}$ and $J_d^{\LMTO}$ numbers can only be used in a comparative way and can not be transferred to LSDA+$U$ within FPLO or to the Hubbard model treatment. For the proper comparison, we perform a constrained LDA calculation for a similar copper oxide material Li$_2$CuO$_2$.\cite{li2cuo2-2009} We find $U_d^{\LMTO}=9.4$~eV and $J_d^{\LMTO}=1.0$~eV. The $J_d^{\LMTO}$ value does not depend on the ligand, while $U_d^{\LMTO}$ for Li$_2$CuO$_2$ is strongly enhanced compared to $U_d^{\LMTO}=6.6$~eV for CuNCN. This analysis suggests that the on-site Coulomb repulsion is ligand-dependent. Nitrogen $p$ orbitals show larger overlap with Cu $d$ orbitals, thus leading to improved screening of the on-site Coulomb repulsion. Thus, nitride materials will generally require lower $U$ values compared to the respective oxides.

After establishing this trend, we can turn to our experience of band structure calculations for copper oxides. For copper oxides treated within FPLO, the $U_d$ values in the range from 6.0 to 8.0~eV are now well established.\cite{drechsler2007,johannes2006,janson2007,janson2008} The comparison of $U_d^{\LMTO}$ for CuNCN and Li$_2$CuO$_2$ (6.6~eV vs. 9.2~eV) suggests lower $U_d$ of $5-6$~eV for CuNCN in FPLO. For $J_d$, we fix the value of 1.0~eV. In the case of the one-orbital Hubbard model, the $U_{\eff}$ values of $3.5-4.5$~eV have been obtained for copper oxides by fitting model simulations to different experimental results (exchange couplings, optical conductivity, electron energy loss spectra, etc.)\cite{helge-m2cuo3,johannes2006,janson2007} For the nitrides, it is reasonable to take a somewhat lower $U_{\eff}$, and we use $U_{\eff}=3.5$~eV in the present analysis. Further experimental studies of CuNCN should provide a more accurate estimate of this parameter.

\section{Results}
\label{results}
\subsection{LDA-based analysis}
\begin{figure}
\includegraphics{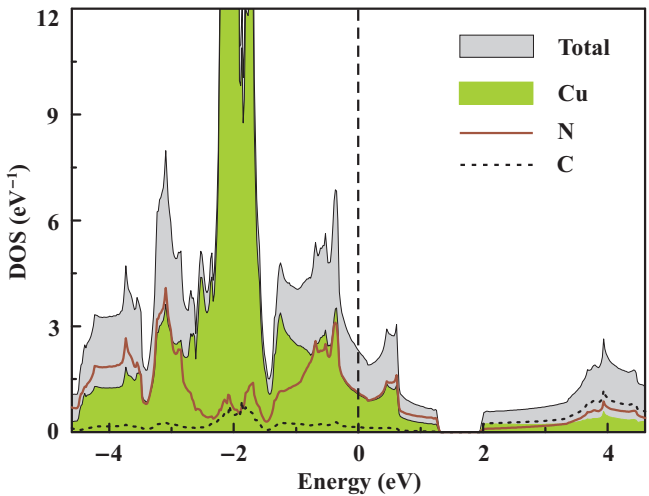}
\caption{\label{dos}
(Color online) LDA density of states for CuNCN. The Fermi level is at zero energy. 
}
\end{figure}

We start with the analysis of LDA band structure of CuNCN. The density of states plot is shown in Fig.~\ref{dos}. Valence bands of CuNCN are formed by Cu $3d$ and N $2p$ orbitals with the minor contribution of carbon states. Empty bands above 2~eV correspond to antibonding states of the NCN groups. The energy spectrum is metallic, because LDA does not account for correlation effects and underestimates band gaps in transition-metal compounds. The energy gap can be reproduced in LSDA+$U$ (see below). 

\begin{figure}
\includegraphics{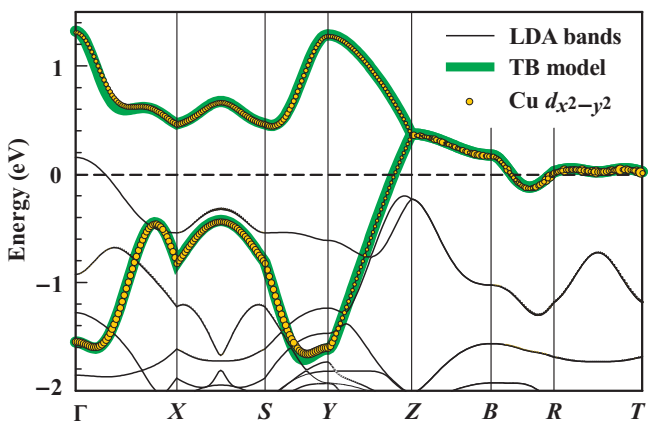}
\caption{\label{bands}
(Color online) LDA bands (thin black lines) and the fit of the TB model (thick green lines). Dots show the contribution of the Cu $d_{x^2-y^2}$ orbital. The notation of $k$ points is as follows: $\Gamma(0,0,0)$, $X(x,0,0)$, $S(\frac{x}{2},0.5,0)$, $Y(0,0.5,0)$, $Z(0,0,0.5)$, $B(0,0.5,0.5)$, $R(\frac{x}{2},0.5,0.5)$, and $T(x,0,0.5)$, where $x=0.25+a^2/4b^2\simeq 0.3087$, and the coordinates are given along $k_x,k_y$, and $k_z$ in units of the respective reciprocal lattice parameters $4\pi/a,4\pi/b$, and $2\pi/c$.
}
\end{figure}

In CuNCN, the local environment of copper is subject to a strong Jahn-Teller distortion. We find four short Cu--N distances of 2.00~\r A within the CuN$_4$ plaquettes and two longer Cu--N distances of 2.61~\r A. Similar to oxides, this type of the local environment should lead to a pronounced crystal field splitting with the highest-lying orbital having $x^2-y^2$ character ($x$ axis runs along one of the short Cu--N bonds, while the $z$ axis is perpendicular to the CuN$_4$ plaquette).\cite{johannes2006,janson2007,janson2008} The respective orbital character is easily recognized in the LDA bands near the Fermi level (Fig.~\ref{bands}). We find two $d_{x^2-y^2}$ bands corresponding to two Cu atoms in the primitive cell of CuNCN. The hybridization to other bands is weak. Therefore, the fit with an effective one-orbital TB model is straightforward. To extract hopping parameters, we calculate overlap integrals of maximally localized Wannier functions centered on Cu sites.\cite{wannier,wannier-fplo} The resulting hoppings $t_i$ are listed in Table~\ref{hoppings}.

\begin{figure}
\includegraphics{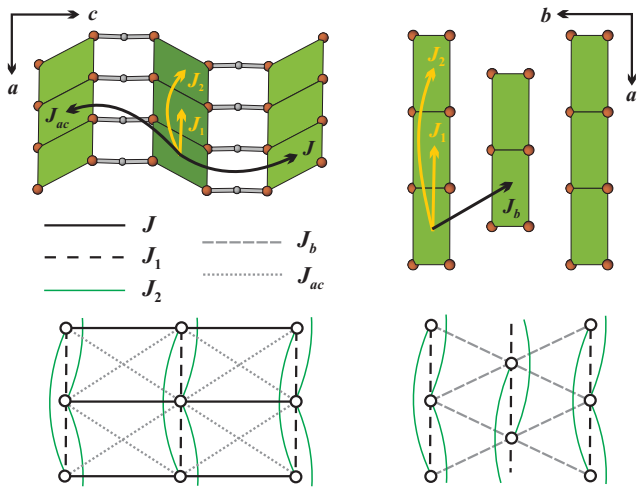}
\caption{\label{interactions}
(Color online) Magnetic interactions in the $ac$ (left) and $ab$ (right) planes of the CuNCN structure. Bottom figures sketch the respective projections of the spin lattice.
}
\end{figure}

The leading hopping runs along the $c$ direction between the structural chains. This hopping $t\simeq -0.384$~eV is very large and leads to the sizable bandwidth $W\simeq 3$~eV. Still, the largest $t$ is one order of magnitude smaller than $U_{\eff}$, hence the perturbation treatment of the corresponding Hubbard model should be reasonable. For the case of the half-filling and for the low-lying excitations, the effective one-orbital Hubbard model is reduced to a Heisenberg model with AFM exchange $J_i^{\AFM}=4t_i^2/U_{\eff}$. Using this simple expression and $U_{\eff}=3.5$~eV, we find the AFM contributions to the exchange couplings, which are listed in Table~\ref{hoppings}. The strongest AFM coupling $J\simeq 1970$~K is mediated by NCN groups (see Fig.~\ref{interactions}). The NNN coupling within the structural chains ($J_2$) and the long-range coupling in the $ac$ plane ($J_{ac1}$) are much weaker and amount to $50-70$~K only. Further hoppings are below 0.03~eV, i.e., the respective $J^{\AFM}$ do not exceed 10~K. In particular, the AFM interaction along the $b$ direction is negligible ($t_b=0.009$~eV, i.e., $J_b^{\AFM}\simeq 1$~K).

\begin{table}
\caption{\label{hoppings}
  Leading hoppings of the TB model $t_i$ (in~eV) and the resulting AFM contributions to the exchange integrals $J_i^{\AFM}=4t_i^2/U_{\eff}$ (in~K), $U_{\eff}=3.5$~eV.
}
\begin{ruledtabular}
\begin{tabular}{cccc}
     $t$     &    $t_1$     &    $t_2$     &     $t_{ac}$    \\
  $-0.384$   &   $0.031$    &   $0.058$    &     $-0.073$    \\\hline
  $J^{\AFM}$ & $J_1^{\AFM}$ & $J_2^{\AFM}$ & $J_{ac}^{\AFM}$ \\ 
   $1960$    &    $13$      &    $45$      &       $71$      \\
\end{tabular}
\end{ruledtabular}
\end{table}

\begin{figure*}
\includegraphics{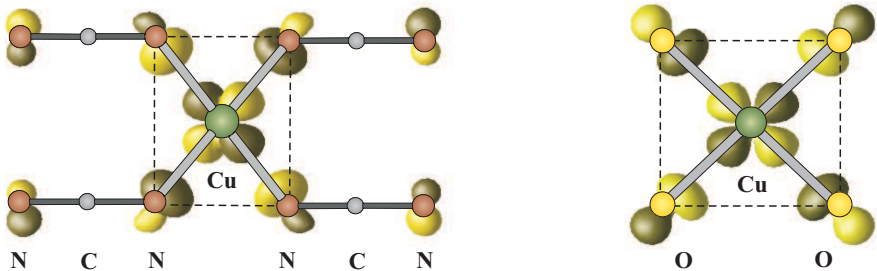}
\caption{\label{wannier}
Maximally localized Wannier functions for CuNCN (left panel) and Li$_2$CuO$_2$ (right panel). The dashed line shows the CuN$_4$ (CuO$_4$) plaquette.
}
\end{figure*}

It is also instructive to consider the shape of the Wannier functions. In Fig.~\ref{wannier}, we plot Wannier functions for CuNCN and for the typical "edge-sharing" oxide material Li$_2$CuO$_2$.\cite{li2cuo2-2009} Each of the Wannier functions is composed of atomic Cu $3d_{x^2-y^2}$ orbital and the $2p$ orbitals of the neighboring nitrogen/oxygen atoms. However, the Wannier function for CuNCN has additional contributions from second-neighbor nitrogen atoms in the NCN groups. The reason for this difference is the strong $\pi$-bonding within the NCN group. In oxides, Cu orbitals usually overlap with individual (atomic) oxygen orbitals. In CuNCN, the Cu $d_{x^2-y^2}$ orbital overlaps with the highest occupied molecular orbitals (HOMO's) of the NCN groups. For the idealized linear NCN group, the HOMO is composed of the two nitrogen orbitals and has $\pi_g$ symmetry.\cite{hoffmann1999} Then, this HOMO enters the Cu-based Wannier function and brings significant contribution from the second-neighbor nitrogen atoms. The spatial extension of the Wannier functions illustrates the large hopping along the $c$ direction and explains the leading magnetic interactions via the NCN groups.

The one-orbital Hubbard model yields AFM couplings only. To estimate FM contributions, one has to extend the model and to include nitrogen $p$ states. The largest FM contribution is expected for $J_1$ due to the $90^{\circ}$ Cu--N--Cu superexchange. For this scenario of edge-sharing copper plaquettes, Mazurenko \textit{et al}.\cite{mazurenko2007} have recently proposed a multi-orbital expression for the total exchange. In a nutshell, their result for the exchange coupling between sites $i$ and $j$ can be written as follows
\begin{equation}
  J_{ij}=\dfrac{4t_{ij}^2}{U_{\eff}}-2\beta^4J_pN_l,
\label{j-fm}\end{equation}
where $\beta$ shows the contribution of each ligand to the correlated bands/Wannier functions, $J_p$ is Hund's coupling at the ligand site, and $N_l$ is the number of ligands where the Wannier functions of adjacent copper sites overlap. Compared to the result of Ref.~\onlinecite{mazurenko2007}, we made two adjustments of the Eq.~\eqref{j-fm}. First, we consider $J_{ij}$ as the full exchange coupling for the $i-j$ bond, hence both terms in Eq.~\eqref{j-fm} are multiplied by a factor of 2. Second, we put $U_{\eff}$ in the denominator of the first term (see footnote for further details).\cite{note1} These adjustments provide transparent physical meaning of the Eq.~\eqref{j-fm}. The first term is the usual AFM coupling resulting from the one-orbital Hubbard model. The second term is the energy gain for the FM configuration due to the Hund's coupling on the ligand site. 

To find $\beta$, we consider the contributions of nitrogen atomic orbitals to the maximally localized Wannier functions centered on Cu sites.
Each nitrogen atom in the CuN$_4$ square shows $\beta^2\simeq 0.088$. Assuming the reasonable value of $J_p=1.5$~eV,\cite{mazurenko2007} we find $J_1^{\FM}\simeq -550$~K. Thus, there is a strong FM coupling along the structural chains. In the following, we will confirm this result using LSDA+$U$ calculations.

\subsection{LSDA+$U$}
According to Sec.~\ref{methods}, $U_d=5-6$~eV is a reasonable range for the LSDA+$U$ Coulomb repulsion parameter in CuNCN. Such $U_d$ values are comparable to the LDA bandwidth $W\simeq 3$~eV, hence one can consider CuNCN as a material on the border between the regimes of strong and moderate correlations. To perform correct modeling, one can approach the problem from both sides: either perform LSDA+$U$ calculations assuming that the correlations are sufficiently strong or solve the multi-orbital Hubbard model using dynamical mean-field theory (DMFT) approach. In the following, we consider the first method and leave a DMFT treatment for future investigations. The reasons for this choice are twofold. First, it is known that LSDA+$U$ removes the quasiparticle band and severely fails in describing spectral functions and metal-insulator transitions. However, LSDA+$U$ should still work for the phenomena related to localized electrons -- e.g., for magnetic interactions in the insulating material CuNCN. Second, weak exchange couplings can be evaluated at sufficiently low temperatures only, while such temperatures are still hardly reachable in the widely-used Monte-Carlo solvers of the DMFT impurity problem. Therefore, we restrict ourselves to the LSDA+$U$ analysis. We will show that LSDA+$U$ yields reasonable results and seems to work quite well even approaching the regime of moderate correlations.

The proximity of $U_d$ to the LDA bandwidth $W$ leads to small energy gaps ($E_g$) and even impedes the convergence of the calculations once the gap approaches zero. This problem is strongly dependent on the double-counting correction and on the spin configuration. We will consider both issues in more detail. The double-counting correction is a necessary part of the LSDA+$U$ approach, because the correlation energy is partially included in LSDA and has to be subtracted, once the explicit Hubbard-like term is added to the energy functional. The two main approaches to the double-counting correction scheme are around-the-mean-field (AMF)\cite{amf} and the fully-localized-limit (FLL).\cite{amf,fll} In the AMF scheme, one derives the double-counting term assuming averaged occupancies for all the correlated orbitals. In contrast, the FLL approach uses integer (0 or 1) occupancies. It is generally believed that the AMF scheme should be used for materials with moderate electronic correlations, while the FLL double-counting correction is more appropriate for strongly correlated (fully localized) systems.\cite{petukhov2003} However, the role of the double-counting correction has been scarcely studied. Recent calculations for model systems established the strong underestimate of the magnetic moment on rare-earth atoms in AMF due to the lack of the Hund's coupling on the correlated site.\cite{ylvisaker} This, however, is irrelevant to the present situation, because copper has one unpaired electron only, and any type of the LSDA+$U$ functional will generally favor the magnetic rather than the non-magnetic state.

For the AMF double-counting correction, spin configurations with AFM ordering along the $c$ axis show lowest energies, largest $E_g$ values, and can be stabilized down to $U_d=3$~eV. For the alternating FM--AFM ordering along the $c$ axis, the convergence is reached at $U_d\geq 5$~eV only. Finally, the FM ordering along the $c$ axis leads to the highest energy and reaches the gapped ground state at $U_d\geq 7$~eV only. To evaluate $J$, one has to use at least two of these configurations. Therefore, the natural lower limit for calculating exchange integrals in CuNCN is $U_d=5$~eV. In the following, we will also consider higher $U_d$ values for the sake of comparison.

In CuNCN, the AMF double-counting correction scheme shows superior results compared to FLL, because the latter yields smaller energy gaps and fails to converge the sufficient set of different spin configurations for $U_d<6$~eV. For $U_d\geq 6$~eV, FLL and AMF yield similar results with a nearly constant offset of $1.5-2.0$~eV in $U_d$ (to reproduce the AMF results with FLL, a higher $U_d$ value should be used). This means that the resulting exchange couplings are robust with respect to the LSDA+$U$ implementation, although the application of FLL requires additional caution. We should note that the application of LSDA+$U$ to moderately correlated systems has already been tested by Petukhov \textit{et al}.\cite{petukhov2003} They used LSDA+$U$ for calculating the band structure of the correlated metal FeAl, found better results for AMF, and suggested the interpolating scheme between the AMF and FLL double-counting corrections. In the present case, this scheme would likely lead to a mere shift of $U_d$, because AMF and FLL yield similar results.

\begin{table}
\caption{\label{lsda+u}
  LSDA+$U$ (AMF) estimates of the exchange integrals (in K), $U_d$ is the Coulomb repulsion parameter (in eV). The notation of the exchange couplings is shown in Fig.~\ref{interactions}. Further couplings are below 5~K.
}
\begin{ruledtabular}
\begin{tabular}{cccccc}
  $U_d$ & $J$  &  $J_1$ & $J_2$ & $J_{ac}$ & $J_b$ \\\hline
    5   & 2900 & $-500$ &  120  &    32    &   4   \\
    6   & 2340 & $-410$ &  98   &    52    &   2   \\
    7   & 1880 & $-330$ &  77   &    70    &   2   \\
    9   & 1210 & $-180$ &  46   &    65    &   2   \\
\end{tabular}
\end{ruledtabular}
\end{table}

In Table~\ref{lsda+u}, we summarize the exchange couplings obtained from AMF LSDA+$U$ calculations. Some of the couplings were estimated in two different supercells (see Sec.~\ref{methods}), and different results were found. For example, at $U_d=5$~eV the calculations for the $2\mathbf{a}\times \mathbf{b}\times \mathbf{c}$ supercell yield $J_1=-544$~K and $J_b=4$~K, while the calculations for the $4\mathbf{a}\times\frac12(\mathbf a+\mathbf b)\times\mathbf c$ supercell lead to $J_1+J_b=-454$~K. The $J_1=-500$~K value in Table~\ref{lsda+u} is a rough average of these estimates. In a similar way, we find 10~\% error bar for $J_{ac}$. Upon increasing $U_d$, the difference between the two supercells is reduced. At $U_d=9$~eV, both supercells show consistent estimates of all the exchange couplings similar to strongly localized systems (e.g., Ref.~\onlinecite{cucl}). This feature may be related to the regime of moderate correlations in CuNCN.\cite{note3} It is also worth to note that the change in the exchange-correlation potential (GGA instead of LDA for one of the supercells) leads to a similar 10~\% error bar for $J$. This error bar persists even at high $U_d$. 

The above considerations show that the LSDA+$U$ estimates of the exchange couplings include the sizable error bar of about 10~\%. Nevertheless, they yield a robust microscopic scenario which is consistent with the LDA-based analysis. The leading coupling $J$ runs along the $c$ axis and reaches nearly 3000~K assuming $U_d=5$~eV. The NN coupling within the structural chains is indeed FM: $J_1\simeq -500$~K. There is also the NNN AFM coupling within the structural chains ($J_2\simeq 100$~K) and the AFM coupling in the $ac$ plane $J_{ac1}=30-50$~K. Further couplings are very weak in agreement with the results of the TB analysis. At higher $U_d$, the scenario persists, but the absolute values of the leading exchange couplings are reduced.
\medskip

Our model analysis and our LSDA+$U$ calculations consistently point to the leading exchange couplings along the $c$ direction in CuNCN. The visualization of the Wannier functions suggests the crucial role of the NCN groups in mediating these interactions. We also find the sizable FM coupling within the structural chains and weaker AFM couplings. These results will be compared to the experimental findings in the next section. In the remainder of this section, we will try to find out why our 1D model with the leading coupling along the $c$ direction is different from the previously proposed 2D model with leading couplings in the $ab$ plane.\cite{cuncn-2008}

While our study and the previous investigation by Liu \textit{et al}.\cite{cuncn-2008} use a similar approach to evaluate the exchange couplings (mapping LSDA+$U$ total energies onto the classical Heisenberg model), there are several differences which might be responsible for the difference in the resulting scenario. These differences are: i) exchange-correlation potential (LDA vs. GGA); ii) basis set (local orbitals in FPLO vs. PAW in VASP); iii) double-counting correction schemes (conventional AMF and FLL vs. FLL-like Dudarev's approach\cite{dudarev} in VASP). The first issue has been discussed above: LDA and GGA yield similar results within FPLO. To test the two other issues, we performed comparative calculations in VASP using both the conventional FLL and the Dudarev's schemes for the double-counting correction (in VASP, the AMF scheme is not available). At $U_d=10$~eV and $J_d=1$~eV (this should roughly correspond to $U_d=9$~eV and $J_d=1$~eV in FPLO), we arrived to the FPLO results within the same 10~\% error bar. However, the decrease of $U_d$ down to 7~eV (corresponding to $U_d-J_d=6$~eV in Ref.~\onlinecite{cuncn-2008} or to $U_d=6$~eV in FPLO) dramatically worsened the convergence. This, however, is hard to detect, because VASP normally achieves the convergence in total energy and does not try to reach good charge convergence. At $U_d=7$~eV, some of the spin configurations lacked the energy gap and failed to converge the charge below $10^{-2}$. Then, the reliability of the resulting total energies might be questionable. Moreover, the metallic regime does not allow to map the energies onto the Heisenberg model. Thus, we can suggest that for moderately correlated systems VASP should be used with caution. In this situation, band structure codes implementing the AMF version of LSDA+$U$ are preferable.

\section{Comparison to the experiment and Discussion}
According to our estimates of the exchange couplings (see Tables~\ref{hoppings} and~\ref{lsda+u}), the uniform spin chain should be a reasonable first approximation to the spin model of CuNCN. The magnetic susceptibility of the uniform spin-$\frac12$ chain shows a maximum at $T_{\max}=T/J\simeq 0.6$ with the reduced susceptibility $\chi^*=0.1469$, where $\chi^*=\chi J/(N_Ag^2\mu_B^2)$, $N_A$ is Avogadro's number, $g$ is the $g$-factor, and $\mu_B$ is Bohr magneton.\cite{johnston2000} Assuming $J\simeq 2500$~K\cite{note4} and $g=2$, we find $T_{\max}\simeq 1500$~K which is well above the experimentally studied temperature range between 1.85~K and 320~K.\cite{cuncn-2008} Below $T_{\max}$, the susceptibility decreases down to the finite value $\chi_0^*\simeq 0.102$ at $T=0$ (at zero temperature, the susceptibility remains finite due to the strong quantum fluctuations in a 1D system).\cite{johnston2000} The range of $1.85-320$~K corresponds to $0\leq T/J\leq 0.15$, where temperature dependence of the susceptibility is weak. Indeed, the experimental data show nearly temperature-independent susceptibility of about $9\cdot 10^{-5}$~emu/mol.\cite{cuncn-2008} This value is in good agreement with $\chi_0\simeq 6.1\cdot 10^{-5}$~emu/mol for the uniform spin-$\frac12$ chain with $J\simeq 2500$~K. The discrepancy can be attributed to additional temperature-independent contributions (core diamagnetism, Van Vleck paramagnetism) and to the possible impurity contribution in Ref.~\onlinecite{cuncn-2008}. To get an accurate experimental estimate of $J$, one has to measure the susceptibility up to $T_{\max}$ (or, at least, sufficiently close to $T_{\max}$). 

1D spin systems do not undergo long-range ordering down to zero temperatures, but non-negligible interchain couplings usually cause magnetic ordering in real low-dimensional materials. In the case of CuNCN, we find a sizable interchain interaction $J_1$ that favors FM ordering pattern along the $a$ direction. Such a pattern is further stabilized by $J_{ac}$ but destabilized by $J_2$ (see Fig.~\ref{interactions}). To get an idea about the resulting magnetic structure, it is instructive to consider the couplings along the $a$ direction within the frustrated spin chain model. In this model, the ordering is FM at $J_2/J_1>\alpha_c=-0.25$, while at lower $J_2/J_1$ values the spiral ordering is stabilized.\cite{zinke2009} In CuNCN, $J_1\simeq -500$~K and $J_2\simeq 100$~K, i.e., $J_2/J_1\simeq -0.2$ is close to $\alpha_c$. On the other hand, the $\alpha_c$ value should inevitably be modified by the strong coupling $J>|J_1|,J_2$. This regime has not been studied theoretically. Nevertheless, one can suggest that the coupling $J$ will drive the system away from the spiral ordering, because strong quantum fluctuations in the quasi-1D system will favor the collinear ordering via the "order-from-disorder" mechanism.\cite{order} The relevant experimental example is Li$_2$CuO$_2$, where the $J_2/J_1$ ratio is close to $\alpha_c$, but the magnetic ordering is collinear (with FM alignment of spins along the frustrated spin chains) due to the non-neglibible interchain couplings.\cite{li2cuo2-neutrons,li2cuo2-2009}

Based on the above discussion, we suggest that CuNCN undergoes long-range ordering with antiparallel alignment of spins along the $c$ direction (due to AFM $J$) and parallel alignment of spins along the $a$ direction (due to FM $J_1$). The ordering along the $b$ direction is more difficult to find out. Both LDA and LSDA+$U$ yield $J_{b}<5$~K, i.e., this coupling is at the border of the accuracy of the present analysis. On this energy scale, additional factors (spin anisotropy, dipolar interactions) may be relevant and will influence the magnetic ordering. Still, it is clear that the frustrated triangular arrangement of Cu atoms within the $ab$ plane (right panel of Fig.~\ref{interactions}) does not lead to any sizable frustration, because the coupling $J_{b}$ is very weak.

To get a rough estimate of the ordering temperature ($T_N$), we use theoretical results for coupled spin chains. Unfortunately, there are no established expressions for the case of different couplings along the $a$ and $b$ directions. Therefore, one has to use a simplified model with the unique effective coupling $J_{\perp}$ between the spin chains. The long-range couplings ($J_2,J_{ac}$) have to be neglected. In the following, we assume $J_{\perp}=(|J_1|+J_b)/2\simeq |J_1|/2\simeq 250$~K. Similar approximations have been used for other spin-chain materials with spatially anisotropic interchain couplings and overestimated $T_N$ by a factor of $3-4$.\cite{helge-m2cuo3,kaul2003} The expression by Schulz\cite{schulz1996} leads to $T_N\simeq 730$~K. Then, the actual $T_N$ should be scaled down to $150-250$~K due to the spatial anisotropy of interchain coupling. Furthermore, one can expect an even lower $T_N$ due to the frustration of the spin system. Thus, the reasonable estimate is $100-150$~K which is still detectable for the experimental methods. We suggest that the 70~K anomaly in the magnetic susceptibility\cite{helge-m2cuo3,cuncn-2008} should be taken as the signature of the long-range ordering in CuNCN. The lack of the respective specific heat anomaly and the lack of magnetic reflections in neutron diffraction can be explained by the pronounced one-dimensionality of the spin system.

The magnitude of the magnetic neutron scattering depends on the ordered moment (sublattice magnetization) of the material. This quantity can be estimated within the two-dimensional (2D) $J-J_1$ model, because 2D systems have finite sublattice magnetization. Quantum Monte-Carlo simulations suggest the ordered moment $\mu\simeq 0.5$~$\mu_B$ for $|J_1|/J=0.2$.\cite{sandvik1999} The reduction in the ordered moment (compared to the classical value of 1~$\mu_B$) is caused by the strong quantum fluctuations arising in the 1D spin system. For other spin-chain materials, even lower $\mu$ values have been reported.\cite{m2cuo3-neutrons} The reduced value of the ordered moment suggests that magnetic reflections in neutron diffraction patterns should be weak. Since the proposed magnetic ordering pattern does not increase the unit cell (and even retains the $C$-centering symmetry in the case of FM spin alignment along the $b$ direction), magnetic reflections should overlap with the nuclear ones, making the experimental observation of the magnetic ordering difficult. This may explain why the experimental neutron diffraction study did not indicate magnetic ordering,\cite{cuncn-2008} while our model suggests the rather high N\'eel temperature of $100-150$~K. To separate magnetic and nuclear reflections, one can apply the polarized neutron scattering technique that has proven to be a sensitive tool for studying long-range ordering in frustrated low-dimensional spin systems with a strongly reduced sublattice magnetization.\cite{skoulatos2009}

The specific heat anomaly arises from the release of entropy upon the transition to the paramagnetic state. If the transition temperature is sufficiently small ($T_N/J\ll 1$), the available entropy is also small, and the anomaly will be completely suppressed. This effect has been studied theoretically for square lattices with a weak interlayer coupling\cite{sengupta2003} and was further confirmed by the experiments on Cu-based square lattice compounds.\cite{lancaster2007} For the uniform spin chain, a similar behavior can be expected. Assuming $T_N/J\simeq 0.030$, we find $S\simeq 0.021R$,\cite{johnston2000} i.e., about 3~\% of the full entropy $R\ln 2$ for a spin-$\frac12$ system or below 2~\% of the lattice entropy at $T_N$ (see Fig.~5 in Ref.~\onlinecite{cuncn-2008}). Since powder samples usually show broad transition anomalies (see, e.g., Ref.~\onlinecite{nath2008}), we believe that the magnetic ordering in CuNCN should lead to a very weak, hardly resolvable specific heat anomaly at $T_N$. To check this hypothesis, specific heat measurements on a single crystal are desirable.

The above discussion shows that our spin model of CuNCN is consistent with all the experimental data available so far. The 70~K susceptibility anomaly is likely an indication of the magnetic ordering with a low ordered moment that impedes the observation of magnetic scattering in conventional neutron diffraction. To study the magnetic ordering in CuNCN, one has to use more sensitive techniques, such as polarized neutron scattering or muon spin relaxation ($\mu$SR). Resonance techniques can also be helpful, because they also evidence the long-range ordering by the shift of the absorption line in electron spin resonance or by the anomaly in the spin-lattice relaxation rate, measured by nuclear magnetic resonance. 

The last comment regarding the experimental data deals with the electronic band gap $E_g$. LSDA+$U$ yields $E_g=1.0$~eV for the ground-state spin configuration at $U_d=5$~eV. This result is in qualitative agreement with the experimentally observed black color of CuNCN, although the resistivity measurements yield a lower activation energy $E_a\simeq 0.1$~eV.\cite{cuncn-2008} However, one should be aware that resistivity measurements yield activation energy for the transport of charge carriers, while this energy is generally unrelated to the electronic band gap (see Ref.~\onlinecite{lifepo4} for an instructive example). In insulating transition metal compounds, charge is usually carried by polarons. Then, $E_a$ is the lattice reorganization energy rather than the gap in the electronic spectrum. To obtain an experimental estimate of $E_g$, optical measurements are necessary.

Now, we will compare CuNCN to copper oxides with similar "edge-sharing" chains of CuX$_4$ squares (X = N, O). Our results show that the introduction of the nitrogen ligand preserves the physics within the structural [CuX$_2$] chains. The unpaired electron occupies the $x^2-y^2$ orbital lying within the CuX$_4$ plaquette. Then, frustrating NN and NNN interactions emerge. The NNN coupling $J_2\simeq 100$~K is comparable to NNN couplings of $60-100$~K in LiCu$_2$O$_2$, LiCuVO$_4$, and other "edge-sharing" chain compounds. In contrast, the NN coupling $J_1\simeq -500$~K is enhanced (the typical values for oxides are $|J_1|\leq 200$~K).\cite{licu2o2-2004,masuda2005,enderle2005,li2cuo2-2009} This effect can be explained by the larger spatial extension of the Wannier functions due to the stronger hybridization between Cu and N orbitals and due to the large contribution of the HOMO of the NCN group. Then, Hund's coupling on the nitrogen site leads to the strong FM interaction, as expressed by the Eq.~\eqref{j-fm}.

However, the above-mentioned similarities do not lead to the similar physics. In CuNCN, the $\pi$-conjugated NCN groups mediate the very strong superexchange coupling $J$ that largely exceeds the couplings within the structural chains. Then, one again finds a 1D magnetic behavior, but the spin chains run perpendicular to the structural chains, and the magnetic behavior resembles the "corner-sharing" scenario of Sr$_2$CuO$_3$ rather than the "edge-sharing" scenario of Li$_2$CuO$_2$, LiCu$_2$O$_2$, and LiCuVO$_4$. This finding reminds different alignment of structural and magnetic chains in vanadium oxides (e.g., Sr$_2$V$_3$O$_9$\cite{kaul2003} or (VO)$_2$P$_2$O$_7$\cite{vop2o7}), although the reason is different. In vanadium compounds, it is possible to suggest the correct spin model by a qualitative analysis based on the location of the magnetic orbital. In CuNCN, the in-plaquette orbital favors both $J$, $J_1$, $J_2$, and $J_{ac}$ (see Fig.~\ref{interactions}). Then, a careful microscopic analysis is necessary to find out the leading interactions and the correct spin model.

Despite this important difference between CuNCN and "edge-sharing" copper oxides, copper cyanodiimide is an interesting compound on its own. The very strong AFM coupling $J\simeq 2000$~K along one direction is comparable to Sr$_2$CuO$_3$, only, and should lead to efficient heat transport within the spin chains. Additionally, this strong coupling leads to the sizable bandwidth $W\simeq 3$~eV. Such a bandwidth is still too small to compete with the on-site Coulomb repulsion. However, the application of high pressure can increase $W$ and drive the system towards a metal-insulator transition and further unusual properties. Experimental studies of these phenomena would be very interesting.

\section{Conclusions}
We have shown that copper cyanodiimide CuNCN should be considered as a uniform spin-$\frac12$ chain system with spin chains running along the $c$ direction. The interchain couplings are found in the $ab$ plane and show frustration within the structural [CuN$_2$] chains. However, the leading interchain coupling $J_1\simeq -500$~K is sufficiently strong and will likely lead to a FM spin alignment along the $a$ direction. We argue that the 70~K anomaly in the magnetic susceptibility can be assigned to the long-range magnetic ordering. The 1D nature of the spin system leads to strong quantum fluctuations that reduce the ordered moment down to $0.5$~$\mu_B$ and can impede the observation of the long-range ordering in conventional neutron diffraction and in the specific heat. Further studies of CuNCN should include an experimental characterization of the magnetic ground state with polarized neutron scattering, $\mu$SR, and resonance techniques. Such studies are currently underway or in preparation. Experimental access to the electronic structure of CuNCN should be possible via optical measurements and photoemission spectroscopy. The potential application of this material could be the ballistic heat transport, while the possible metal-insulator transition under high pressure will be of fundamental interest.

\acknowledgments
We would like to acknowledge Vladimir Mazurenko and Vladimir Anisimov for providing the modified version of the TB-LMTO-ASA code and Klaus Koepernik for implementing the Wannier functions in FPLO. We are also grateful to Peter H\"ohn for sharing our interest in CuNCN and to Richard Dronskowski for stimulating our work on this compound. A.Ts. acknowledges financial support of MPI PKS.


\begin{thebibliography}{63}
\bibitem[{not({\natexlab{a}})}]{note1}
\bibinfo{note}{The original expression of Ref.~\onlinecite{mazurenko2007}
  includes the AFM part of the exchange in the form
  $$\dfrac{2t^2}{\alpha^4(U_d-V_{dd})+\beta^4J_pN_l/2},$$ where $\alpha$ is the
  contribution of the Cu atomic orbital to the Wannier function and $V_{dd}$ is
  the intersite Coulomb repulsion. This expression results from a multi-orbital
  ($pd$) Hubbard model. The denominator can be considered as the Coulomb
  repulsion in the copper orbital ($U_d$) scaled by the contribution of this
  orbital to the Wannier function/correlated band. Thus, the meaning of the
  parameter in the denominator is similar to that of $U_{\eff}$. Since the
  unambiguous evaluation of the $U_d$ and $V_{dd}$ parameters is difficult, we
  further estimate $J_1^{\AFM}$ as $4t_1^2/U_{\eff}$. In the present case, this
  assumption has little influence on the results, because $t_1$ is very small,
  and $J_1^{\AFM}\ll J_1^{\FM}$.}
\bibitem[{not({\natexlab{b}})}]{note3}
\bibinfo{note}{The low $U_d$ values lead to the reduction in the Cu magnetic
  moment, and the spin polarization gets slightly delocalized. This can enhance
  the long-range couplings that are not included in the model and show
  different energy contributions in different supercells. However, such
  couplings are "dummy" and do not appear in the TB model. Then, they should
  not be included in the resulting spin model as well.}
\end{thebibliography}
\end{document}